\documentclass[reprint,amsmath,amssymb,aps,prb]{revtex4-1}
\usepackage{graphicx}
\usepackage{dcolumn}
\usepackage{bm}
\usepackage{color}
\usepackage{braket}

\newcommand{\tm}[1]{\textcolor{black}{#1}}
\newcommand{\be}{\begin{equation}}
\newcommand{\ee}{\end{equation}}
\newcommand{\bea}{\begin{eqnarray}}
\newcommand{\eea}{\end{eqnarray}}

\begin{document}

\title{Short-Range Antiferromagnetic Correlation Effect
on Conduction Electrons
in Two-Dimensional Strongly Correlated Electron Systems
}

\author{Takao Morinari}
 \email{morinari.takao.5s@kyoto-u.ac.jp}
\affiliation{
  Graduate School of Human and Environmental Studies,
  Kyoto University, Kyoto 606-8501, Japan
}

\date{\today}

\begin{abstract}
We investigate magnetic polarons in two-dimensional
strongly correlated electron systems,
where conduction electrons interact with antiferromagnetically
interacting localized spins.
Starting from a basic model, we derive a simplified model
with the help of spin Green's function and 
a perturbation analysis.
A strong coupling analysis is applied to the model,
where the sum of the scattering wave vectors 
is approximated to be $(\pi,\pi)$ or zero,
using the equation of motion for the conduction electron
Green's function, and we 
discuss the pseudogap like behavior associated with
the suppression of the quasiparticle weights
and the crossover from the large magnetic polaron
to the small magnetic polaron.
In the antiferromagnetic long-range ordered state,
the spectral weight of the conduction electrons
has a form of broad humps due to Franck-Condon broadening
associated with the multi-magnon scattering.
The band folding feature due to the $(\pi,\pi)$ scattering
disappears as we increase the number of the magnons
involved in the multi-magnon scattering.
\end{abstract}

\maketitle

\section{
\label{sec:Intro}
Introduction
}
In the hole doped 
cuprate high-temperature superconductors,\cite{Keimer2015}
one of the key correlations is the antiferromagnetic (AF) correlation,
which is long-ranged in the parent compound but 
short-ranged in moderately doped compounds.
For the purpose of understanding
the physics of the cuprates, especially the enigmatic pseudogap state,
we need to clarify to what extent it is understood
on the basis of the AF correlation.
Although this is less ambitious goal,
it is not an easy task 
because of the strong electronic correlation
which makes the parent compound a charge-transfer insulator.\cite{Zaanen1985}
In Ref.~\onlinecite{Morinari2018}, 
it was pointed out that there is a close relationship between 
the pseudogap and the short-range AF correlation:
The magnetic-torque measurement result,\cite{Sato2017}
whose characteristic temperature
coincides with the pseudogap temperature
determined by other experiments,
has a non-trivial scaling relationship with
the AF spin susceptibility.\cite{Morinari2018}
A pseudogap behavior associated with
the short-range AF correlation
has been discussed in numerical simulations,
such as extended versions of dynamical mean-field theory
\cite{Kyung2006,Ferrero2009,Macridin2006,Sordi2012,Gunnarsson2015}
and quantum Monte Carlo simulations.\cite{Macridin2006,Sordi2012}
In experiments,
the Fermi surface topology changes abruptly 
from arcs\cite{Damascelli2003} to closed contours
in Bi$_2$Sr$_2$CaCu$_2$O$_{8+\delta}$
as observed in scanning tunneling microscopy,\cite{Fujita2014}
where the arcs end at the AF zone boundary.
In angle-resolved photoemission spectroscopy,
the AF zone boundary effect is clearly seen in 
the electron doped cuprate,\cite{Matsui2007,Armitage2010}
while it just gives a terminating point of Fermi arcs 
in the hole doped cuprate.
A possible interpretation is the difference
in the range of the AF
correlation length.\cite{Proust2019}

In this paper we investigate 
the effect of the short-range AF correlation
on conduction electrons.
We focus on low doping systems.
The subject has been studied as a magnetic polaron formed 
in the $t$-$J$ model\cite{Schmitt-Rink1988,Kane1989,Su1989,Martinez1991,Poilblanc1992,Manousakis2007,Grusdt2018,grusdt2019microscopic}
or in the Hubbard model\cite{Brinkman1970,Bulaevski1968,Trugman1988,Valkov2016} 
with the AF correlation.
Although there are powerful numerical simulations mentioned above,
it is useful to study the system in a different way
in order to examine limitations in the numerical simulations 
arising from the momentum resolution.\cite{Gull2010}
Here, we take a strong coupling approach
on the basis of the equation of motion for the Green's function.
We start with a model consisting of conduction electrons
and antiferromagnetically interacting localized spins 
with an exchange coupling between them.
We introduce these degrees of freedom as separate fields
to focus on the interaction effect between them.
From the analysis of the second order perturbation theory,
we find that the coupling between the conduction electrons
and the magnons takes the largest value at 
the scattering wave vector ${\bm Q}=(\pi,\pi)$.
(Hereafter, we take the lattice constant unity.)
On the basis of this observation, we propose a simplified model, 
which is similar to the Holstein model\cite{holstein1959a,holstein1959b} 
for the polaron problem.\cite{Devreese2009}
The difference is just the scattering wave vector.
The advantage of our approach
is that one can control the strength 
of the magnetic correlation effect
by varying the number of magnons
and there is no limitation 
arising from the momentum resolution.
%
%

The rest of the paper is organized as follows.
In Sec.~\ref{sec:Model} we describe our model.
In Sec.~\ref{sec:AFSRO} we examine the short-range AF
correlation.
We propose a simplified model to describe
the system.
A strong coupling analysis is applied to the simplified model.
We derive a general formula to investigate the electron Green's function.
And then, we introduce a dilute magnon approximation applicable
for low-temperatures.
In Sec.~\ref{sec:result} we present the numerical calculation results.
In Sec.~\ref{sec:summary} we summarize the result.

\section{
\label{sec:Model}
Model
}
We consider a strongly correlated two-dimensional electron system
consisting of conduction electrons and localized moments.
We assume that there is a strong exchange interaction
between the conduction electron spins and the localized moments.
The model is taken as the low-energy effective theory
for multi-orbital Hubbard models:
Some of the electrons are localized due to 
a strong on-site Coulomb repulsion,
and form the localized moments.
It is possible to derive a similar model
starting from a single band model,
such as the Hubbard model or the $t$-$J$ model,
by introducing localized moments through 
a Storatonovich-Hubbard transformation
or applying a slave-particle formalism.\cite{Lee2006}
In order to make the situation simple,
we introduce the conduction electrons and the localized moments
as separate fields.

The Hamiltonian is given by
\be
{\mathcal H} = \sum\limits_{{\bm{k}},\sigma } 
{{\xi _{\bm{k}}}c_{{\bm{k}}\sigma }^\dag 
{c_{{\bm{k}}\sigma }}}  
+ \frac{K}{{\sqrt N }}\sum\limits_{{\bm{k}},{\bm{q}}} 
{{{\bm{S}}_{\bm{q}}} \cdot 
\left( {c_{{\bm{k}} + {\bm{q}}}^\dag 
{\bm \sigma} {c_{\bm{k}}}} \right)}  
+ {{\mathcal{H}}_{{\rm{spin}}}},
\label{eq:H}
\ee
where the energy dispersion of the conduction electron, 
${\varepsilon _{\bm{k}}}$, minus the chemical potential, $\mu$, 
is denoted by $\xi _{\bm{k}} = {\varepsilon _{\bm{k}}} - \mu$.
We consider a square lattice, and ${\varepsilon _{\bm{k}}}$ is given by
\bea
{\varepsilon _{\bm{k}}} &=& - 2t\left( {\cos {k_x} + \cos {k_y}} \right) 
- 4{t_1}\cos {k_x}\cos {k_y} \nonumber \\
& & - 2{t_2}\left( {\cos 2{k_x} + \cos 2{k_y}} \right),
\eea
with $t$ the nearest neighbor hopping parameter,
$t_1$ the second nearest neighbor hopping parameter,
and 
$t_2$ the third nearest neighbor hopping parameter.
The creation operator of the conduction electron 
with the wave vector $\bm{k}$ and spin $\sigma$
is denoted by $c_{{\bm{k}}\sigma }^\dag$.
The localized spin moment at site $j$
is denoted by ${\bm S}_j$.
For the value of the spins of the localized moments, 
we assume one-half.
The Fourier transform of ${\bm S}_j$
is denoted by ${\bm S}_{\bm q}$ with $\bm q$ the wave vector.

The second term in the right-hand side of Eq.~(\ref{eq:H}) describes 
the AF exchange coupling between the localized spins
and the conduction electron spins with the coupling constant $K$.
The number of the lattice sites is denoted by $N$.
The conduction electron spin is denoted by using a two-component operator,
$c_{\bm{k}}^\dag  = \left( 
{
{c_{{\bm{k}} \uparrow }^\dag },  {c_{{\bm{k}} \downarrow }^\dag }
} 
\right)$.
The components of the three dimensional vector
${\bm \sigma}  = \left( {{\sigma _x},{\sigma _y},{\sigma _z}} \right)$
are the Pauli matrices.
The last term in Eq.~(\ref{eq:H}) describes 
the interaction between the localized spins.
Here, we consider the AF Heisenberg model
on the square lattice:
\be
{{\mathcal{H}}_{{\rm{spin}}}} 
= J \sum\limits_{\left\langle {i,j} \right\rangle } 
{{{\bm{S}}_i} \cdot {{\bm{S}}_j}},
\ee
where the summation is taken over pairs of nearest neighbor sites
and $J(>0)$ is the exchange interaction.

\section{
\label{sec:AFSRO}
Effect of the AF Short-Range Correlation
on Conduction Electrons
}
\subsection{Antiferromagnetic Short-Range Order}
We may expect that there are various phases,
including the AF long-range ordered phase,
the ferromagnetic metallic phase like 
manganese oxides,\cite{Dagotto2001}
the Fermi liquid phase, etc.,
where the Hamiltonian (\ref{eq:H}) is applied 
with taking a suitable set of parameters.
Here,
we focus on a metallic phase without AF long-range order.
In the presence of the conduction electrons, 
one may expect that
the exchange interaction $J$ is reduced from the original value.
So, $J$ must be replaced by an effective exchange interaction
depending on the concentration of the conduction electrons.
Hereafter, we denote this effective exchange interaction
by the same symbol $J$ to make the notation simple.
In addition, the dynamics of the localized spins can be
affected by the conduction electrons.
To make the situation simple, we assume that 
the number of conduction electrons is small,
and we neglect the effect of the conduction electrons
on the localized spin dynamics.

We are interested in finite temperatures
where a mean field approach, 
for example,
a Schwinger boson mean field theory\cite{Arovas1988} and
a modified spin-wave theory\cite{Takahashi1989},
is not reliable.
These mean field theories provide an accurate description
of the ground state properties,
while they fail to describe features for
$T > T_0$ with $T_0 \sim 0.7 J$,\cite{Yoshioka1989} 
in particular a broad peak\cite{Manousakis1991} 
in the temperature dependence of the spin-susceptibility.
The spin-spin correlation associated with this broad peak 
is described by the Green's function method.\cite{Kondo1972,Shimahara1991,Winterfeldt1997,Zavidonov1998,Sadovskii2001}
So, we use this formulation 
for the description of the localized spins.
The formulation is briefly reviewed in Appendix~\ref{app:AFSRO}.

\subsection{The Second Order Perturbation Theory}
Now we examine the effect of AF short-range order on
the conduction electrons.
Theoretically challenging point is that the conduction electrons
strongly coupled with the localized spins.
Therefore, we need to apply a strong coupling analysis.
\tm{
  In general, we need both higher-order terms
  and some kind of self-consistent calculation
  for any strong coupling analysis.
  However, this is a formidable task
  because rapidly increases
  the number of summations of
  the wave vectors and the Matsubara frequencies.
  So, we need to introduce some approximation.
  For the purpose of carrying out a strong coupling analysis, 
}
we take the following strategy:
First, we apply the standard second order perturbation theory
\cite{mahan2013many}
to the system.
We examine the result and try to extract essential properties.
And then, we construct a simplified model,
to which one can apply a strong coupling analysis
with the help of some approximation.

We consider the Matsubara Green's function for the conduction electron
with wave vector ${\bm k}$ and spin $\sigma$,
which is given by
\be
{G_{{\bm{k}}\sigma }}\left( \tau  \right) 
=  - \left\langle {{T_\tau }{c_{{\bm{k}}\sigma }}\left( \tau  \right)
c_{{\bm{k}}\sigma }^\dag \left( 0 \right)} \right\rangle,
\label{eq:Gks}
\ee
with $\tau$ the imaginary time.
Here, $T_{\tau}$ is the imaginary time ordering operator
and
$c_{{\bm{k}}\sigma }\left( \tau  \right) 
= \exp \left( \tau \mathcal{H} \right)
c_{{\bm{k}}\sigma }
\exp \left(  - \tau \mathcal{H} \right)$
with the Hamiltonian being given by Eq.~(\ref{eq:H}).

We take the second term in Eq.~(\ref{eq:H}) as the perturbation
as if the coupling constant $K$ were a small parameter.
It is easy to find that the first-order electron self-energy vanishes.
The second-order electron self-energy is given by
\be
\Sigma _{{\bm{k}}\sigma }^{\left( 2 \right)}\left( {i{\omega _n}} \right) 
= \frac{{3{K^2}}}{{2\beta N}}\sum\limits_{i{\Omega _n}}
\sum\limits_{\bm{q}} 
D_{\bm{q}} \left( {i{\Omega _n}} \right)
G_{{\bm{k}} + {\bm{q}},\sigma }
\left( {i{\omega _n} + i{\Omega _n}} \right),
\label{eq:Sigma_2}
\ee
where $\beta=1/T$ and the magnon propagator,
which is presented in Appendix \ref{app:AFSRO},
is given by
\be
{D_{\bm{q}}}\left( {i{\Omega _n}} \right) 
=  - \frac{{4J{c_1}\left( {1 - {\gamma _{\bm{q}}}} \right)}}
{{{{\left( {i{\Omega _n}} \right)}^2} - \omega _{\bm{q}}^2}},
\ee
where $\gamma_{\bm q}=(\cos q_x + \cos q_y)/2$
and $c_1$ is the spin-spin correlation between the nearest neighbor sites,
which is defined by $c_1 = 2 \langle S^+_i S^-_j \rangle$
with $i$ and $j$ being nearest neighbor sites.
Here, we set the Boltzmann constant $k_{\rm B}=1$.
Carrying out the summation over the Matsubara frequency 
$\Omega_n = 2\pi n/\beta$ with $n$ an integer,
we obtain
\bea
\Sigma _{{\bm{k}}\sigma }^{\left( 2 \right)}\left( {i{\omega _n}} \right) 
&=&  \frac{{{K^2}}}{N}\sum\limits_{\bm{q}} 
g_{\bm q} \left[ \frac{{n_B\left( {{\omega _q}} \right) 
+ f\left( {{\xi _{{\bm{k}} + {\bm{q}}}}} \right)}}{{i{\omega _n} 
- {\xi _{{\bm{k}} + {\bm{q}}}} + {\omega _q}}} 
\right. \nonumber \\
& & \left. + \frac{{n_B\left( {{\omega _q}} \right) + 1 
- f\left( {{\xi _{{\bm{k}} + {\bm{q}}}}} \right)}}{{i{\omega _n} 
- {\xi _{{\bm{k}} + {\bm{q}}}} - {\omega _q}}} \right],
\label{eq:Sigma_2form}
\eea
with $n_B$ the Bose-Einstein distribution function
and $f$ the Fermi-Dirac distribution function.
The ${\bm q}$ dependent coupling is given by
\be
{g_{\bm{q}}} 
= \frac{{3|c_1| J\left( {1 - {\gamma _q}} \right)}}{{{\omega _q}}}. 
\ee

The expression of 
$\Sigma _{{\bm{k}}\sigma }^{\left( 2 \right)}\left( {i{\omega _n}} \right)$
is a familiar result found in an electron-boson 
coupled system.\cite{mahan2013many}

\subsection{The Simplified Model}
An important point about Eq.~(\ref{eq:Sigma_2form})
is that the right-hand side is independent of the spin of
the conduction electron.
This is distinct from the case with the AF long-range ordered state
where we need to study each spin state separately.
\cite{Schmitt-Rink1988,Kane1989,Su1989,Martinez1991}
Another important point is that there is no need to
distinguish even sites and odd sites.
In the presence of the AF long-range order,
we need to distinguish them separately.
In addition, the wave vector takes 
the values in the full Brillouin zone,
and not restricted to the reduced 
magnetic Brillouin zone.

\begin{figure}[htbp]
\includegraphics[width=0.5\textwidth]{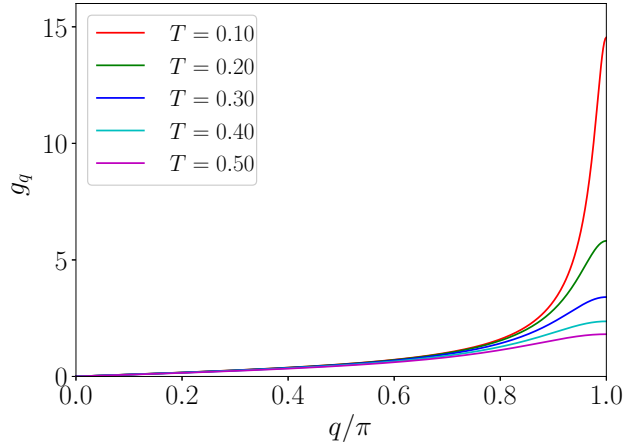}
\caption{
  \label{fig:g_q} 
  (Color online)
  The wave vector ${\bm q}=(q,q)$ dependence of the coupling $g_{\bm q}$
  in Eq.~(\ref{eq:Sigma_2form})
  for different temperatures.
  The coupling $g_{\bm q}$ takes the maximum
  at ${\bm q}=(\pi,\pi)$,
  and the maximum value increases as the temperature is lowered.
}
\end{figure}

In Eq.~(\ref{eq:Sigma_2form}), 
the information about the short-range AF order
is included through the magnon dispersion $\omega_{\bm q}$
and the ${\bm q}$-dependent coupling $g_{\bm q}$.
The self-energy (\ref{eq:Sigma_2form}) itself has a standard form
where conduction electrons couple with
bosonic excitations.\cite{mahan2013many}
In Fig.~\ref{fig:g_q}, we show ${\bm q}$ dependence of $g_{\bm q}$
for different temperatures.
The crucial point here is that $g_{\bm q}$
exhibits a sharp peak at ${\bm q}=(\pi,\pi)$.

From the consideration above, 
we consider a simplified model
with the characteristic features
of $g_{\bm q}$ and $\omega_{\bm q}$,
that is,
\bea
{\mathcal H} &=& \sum\limits_{\bm{k}} 
{{\xi _{\bm{k}}}c_{\bm{k}}^\dag {c_{\bm{k}}}}  
+ \frac{g}{{\sqrt N }}\sum\limits_{{\bm{k}},{\bm{q}}} 
{\left( {b_{\bm{q}}^\dag  
+ {b_{ - {\bm{q}}}}} \right)c_{\bm{k}}^\dag 
{c_{{\bm{k}} + {\bm{q}} + {\bm Q} }} }
\nonumber \\
& & + \sum\limits_{\bm{q}} {\Omega b_{\bm{q}}^\dag {b_{\bm{q}}}}.
\label{eq:tm_model}
\eea
Here, we omit the spin dependence of the conduction electrons
because there is no need to distinguish the spin-up and spin-down states.
The magnon excitation $\omega_{{\bm Q}+{\bm q}}$
is created (annihilated) by $b_{\bm{q}}^\dag$ ($b_{\bm{q}}$).
We neglect the dispersion of $\omega_{{\bm Q}+{\bm q}}$,
and take $\omega_{{\bm Q}+{\bm q}} \simeq \omega_{\bm Q} \equiv \Omega$ 
as an approximation
because of the behavior of $g_{\bm q}$ as discussed above.
The coupling constant $g$ is chosen as 
the value of $K^2 g_{\bm q}$ at ${\bm q}={\bm Q}$.

In the hole doped 
cuprate high-temperature superconductors,
$\Omega$ is associated with the resonance energy
at the wave vector ${\bm Q}=(\pi,\pi)$
observed in the neutron scattering,
\cite{hayden2004structure,tranquada2004quantum,Fujita2012}
from which broad peaks disperse upward
and incommensurate peaks disperse downward,
resulting in the hourglass pattern.

\subsection{Strong Coupling Analysis}
The model (\ref{eq:tm_model}) is similar to
the Holstein model for the polaron\cite{holstein1959a,holstein1959b}.
The difference is just that the shift of the wave vector ${\bm Q}$
at the scattering of the conduction electrons by the bosons.

Here, we are interested in the strong coupling regime for $g$.
So, we need to apply a strong coupling analysis.
In the study of the Holstein model, various approaches
have been applied in the strong coupling regime.
Among others, the momentum average approximation
\cite{Berciu2006,Goodvin2006}
is a useful approach which reproduces 
the most reliable diagrammatic Monte Carlo results.
\cite{Prokofiev1998,Mishchenko2000}
Here, we apply a modified version 
of the momentum average approximation
to the model (\ref{eq:tm_model}).

Now we assume that the carrier density is low enough 
so that we may consider a single carrier system.
Omitting the spin, $\sigma$,
the equation of motion 
of the Green's function (\ref{eq:Gks}) 
is given by
\be
i{\omega _n}{G_{\bm{k}}}\left( {i{\omega _n}} \right) 
= {\left\langle {{\left[ {{c_{\bm{k}}},{\mathcal H}} \right]}} 
\left| 
 {{c_{\bm{k}}^\dag }} 
\right.
\right\rangle _{i{\omega _n}}} + 1.
\label{eq:G1}
\ee
Here, the Hamiltonian is given by Eq.~(\ref{eq:tm_model})
and the notation is defined in Appendix \ref{app:AFSRO}.
The commutator in the right-hand side is
\be
\left[ {{c_{\bm{k}}},{\mathcal H}} \right] 
= {\varepsilon _{\bm{k}}}{c_{\bm{k}}} + \frac{g}{{\sqrt N }}
\sum\limits_{\bm{q}} {\left( {b_{\bm{q}}^\dag  
+ {b_{ - {\bm{q}}}}} \right){c_{{\bm{k}} + {\bm{q}} + {\bm{Q}}}}}.
\ee
After substituting this equation into Eq.~(\ref{eq:G1}),
we need to compute,
${\left\langle {{{b_{ - {\bm{q}}}}{c_{{\bm{k}} + {\bm{q}} + {\bm{Q}}}}}}
\Big|
{{c_{\bm{k}}^\dag }} \right\rangle _{i{\omega _n}}}$,
and
${\left\langle 
b_{\bm{q}}^{\dagger}
{c_{{\bm{k}} + {\bm{q}} + {\bm{Q}}}}
\Big|
{{c_{\bm{k}}^\dag }} \right\rangle _{i{\omega _n}}}$.
We consider the equation of motion of these quantities,
and then repeat the similar procedure.
In order to carry out the calculation in a systematic way,
we define
\bea
\lefteqn{
{G_{n,m}}\left( {{\bm{k}},i{\omega _n},
{{\bm{q}}_1},{{\bm{q}}_2},...,{{\bm{q}}_n},
{{\bm{p}}_1},{{\bm{p}}_2},...,{{\bm{p}}_m}} \right)
}
\\
&=& 
\left\langle {{b_{{{\bm{q}}_1}}^\dag 
...b_{{{\bm{q}}_n}}^\dag 
{b_{ - {{\bm{p}}_1}}}
...{b_{ - {{\bm{p}}_m}}}{c_{{\bm{k}} 
+ {\bm{q}}_T^{\left( n \right)} + {\bm{p}}_T^{\left( m \right)}}}}}
\right| \left. {{c_{\bm{k}}^\dag }} \right\rangle _{i{\omega _n}}
\eea
where
\be
{\bm{q}}_T^{\left( n \right)} = 
\sum\limits_{j = 1}^n {{\bm{q}}_j} 
+ n {\bm Q} .
\ee
\begin{widetext}
The equation of motion is,
\bea
\lefteqn{
i{\omega _n}{G_{n,m}}\left( {{\bm{k}},i{\omega _n},{{\bm{q}}_1},{{\bm{q}}_2},...,{{\bm{q}}_n},{{\bm{p}}_1},{{\bm{p}}_2},...,{{\bm{p}}_m}} \right)
}
\\ \nonumber
&=& {\left\langle {{
\left[ {b_{{{\bm{q}}_1}}^\dag b_{{{\bm{q}}_2}}^\dag ...b_{{{\bm{q}}_n}}^\dag {b_{ - {{\bm{p}}_1}}}{b_{ - {{\bm{p}}_2}}}...{b_{ - {{\bm{p}}_m}}}{c_{{\bm{k}} + {\bm{q}}_T^{\left( n \right)} + {\bm{p}}_T^{\left( m \right)}}},H} \right]}}
\right| \left.
 {{c_{\bm{k}}^\dag }} \right\rangle _{i{\omega _n}}} 
\nonumber \\ & & 
+ \left\langle {\left\{ 
{b_{{{\bm{q}}_1}}^\dag 
...b_{{{\bm{q}}_n}}^\dag 
{b_{ - {{\bm{p}}_1}}}
...{b_{ - {{\bm{p}}_m}}}{c_{{\bm{k}} 
+ {\bm{q}}_T^{\left( n \right)} 
+ {\bm{p}}_T^{\left( m \right)}}},c_{\bm{k}}^\dag } \right\}} \right\rangle.
\label{eq:GnmEOM}
\eea
We compute the commutator in the right-hand side.
Noting that there is only a single carrier,
we obtain
\bea
\lefteqn{
\left[ {i{\omega _n} - {\varepsilon _{{\bm{k}} 
+ {\bm{q}}_T^{\left( n \right)} + {\bm{p}}_T^{\left( m \right)}}} 
- \left( {m - n} \right)\Omega } \right]
{G_{n,m}}\left( {{\bm{k}},i{\omega _n},
{{\bm{q}}_1},
...,
{{\bm{q}}_n},
{{\bm{p}}_1},
...,{{\bm{p}}_m}} \right) }
\nonumber
\\
&=& \frac{g}{{\sqrt N }}\sum\limits_{{{\bm{p}}_{m + 1}}} {{G_{n,m + 1}}\left( {{\bm{k}},i{\omega _n},{{\bm{q}}_1},{{\bm{q}}_2},...,{{\bm{q}}_n},{{\bm{p}}_1},{{\bm{p}}_2},...,{{\bm{p}}_{m + 1}}} \right)} 
\nonumber \\ & & 
+ \frac{g}{{\sqrt N }}\sum\limits_{{{\bm{q}}_{n + 1}}} {{G_{n + 1,m}}\left( {{\bm{k}},i{\omega _n},{{\bm{q}}_1},...,{{\bm{q}}_{n + 1}},{{\bm{p}}_1},...,{{\bm{p}}_m}} \right)} 
\nonumber
\\
& & + \frac{g}{{\sqrt N }}\sum\limits_{j = 1}^m {{G_{n,m - 1}}\left( {{\bm{k}},i{\omega _n},{{\bm{q}}_1},...,{{\bm{q}}_n},{{\bm{p}}_1}...,\widehat {{{\bm{p}}_j}},...,{{\bm{p}}_m}} \right)} 
\nonumber \\ & & 
+ {\delta _{n,m}}{\delta _{{\bm{q}}_T^{\left( n \right)} + {\bm{p}}_T^{\left( n \right)},0}}\left\langle {b_{{{\bm{q}}_1}}^\dag b_{{{\bm{q}}_2}}^\dag ...b_{{{\bm{q}}_n}}^\dag {b_{ - {{\bm{p}}_1}}}{b_{ - {{\bm{p}}_2}}}...{b_{ - {{\bm{p}}_n}}}} \right\rangle,
\label{eq:GnmEOM2}
\eea
where $\widehat {\bm{p}}_j$ denotes that
${\bm{p}}_j$ is excluded.
The full Green's function is given by
\be
G\left( {{\bm{k}},i{\omega _n}} \right) 
= {G_{0,0}}\left( {{\bm{k}},i{\omega _n}} \right) 
= G_{\bm{k}}^{\left( 0 \right)}\left( {i{\omega _n}} \right)
\left[ {1 + g\sqrt N {g_{0,1}}\left( {{\bm{k}},i{\omega _n}} \right) + g\sqrt N {g_{1,0}}\left( {{\bm{k}},i{\omega _n}} \right)} \right],
\ee
where
\be
G_{\bm{k}}^{\left( 0 \right)}\left( {i{\omega _n}} \right) 
= \frac{1}{{i{\omega _n} - {\varepsilon _{\bm{k}}}}}.
\ee

Here, ${g_{n,m}}\left( {{\bm{k}},i{\omega _n}} \right)$ is defined by
\bea
{g_{n,m}}\left( {{\bm{k}},i{\omega _n}} \right) 
= \frac{1}{{{N^{n + m}}}}
\sum\limits_{{{\bm{q}}_1},...,{{\bm{q}}_n},{{\bm{p}}_1},...,{{\bm{p}}_m}} 
{{G_{n,m}}\left( {{\bm{k}},i{\omega _n},
{{\bm{q}}_1},...,{{\bm{q}}_n},
{{\bm{p}}_1},...,{{\bm{p}}_m}} \right)}.
\eea
The recursion formula for 
${g_{n,m}}\left( {{\bm{k}},i{\omega _n}} \right)$
is found from Eq.~(\ref{eq:GnmEOM2}), and is given by
\bea
{g_{n,m}}\left( {{\bm{k}},i{\omega _n}} \right) 
&=& \frac{{g\sqrt N }}{{{N^{n + m + 1}}}}\sum\limits_{{{\bm{q}}_1},...,{{\bm{q}}_n},{{\bm{p}}_1},...,{{\bm{p}}_{m + 1}}} {G_{{\bm{k}} + {\bm{q}}_T^{\left( n \right)} + {\bm{p}}_T^{\left( m \right)}}^{\left( 0 \right)}\left( {i{\omega _n} - \left( {m - n} \right)\Omega } \right)} 
\nonumber \\ & & \times 
{G_{n,m + 1}}\left( {{\bm{k}},i{\omega _n},{{\bm{q}}_1},...,{{\bm{q}}_n},{{\bm{p}}_1},...,{{\bm{p}}_{m + 1}}} \right)
\nonumber \\
& & + \frac{{g\sqrt N }}{{{N^{n + m + 1}}}}\sum\limits_{{{\bm{q}}_1},...,{{\bm{q}}_{n + 1}},{{\bm{p}}_1},...,{{\bm{p}}_m}} {G_{{\bm{k}} + {\bm{q}}_T^{\left( n \right)} + {\bm{p}}_T^{\left( m \right)}}^{\left( 0 \right)}\left( {i{\omega _n} - \left( {m - n} \right)\Omega } \right)} 
\nonumber \\ & & \times 
{G_{n + 1,m}}\left( {{\bm{k}},i{\omega _n},{{\bm{q}}_1},...,{{\bm{q}}_n},{{\bm{p}}_1},...,{{\bm{p}}_{m + 1}}} \right)
\nonumber \\
& & + \frac{{g\sqrt N }}{{{N^{n + m}}}}\sum\limits_{{{\bm{q}}_1},...,{{\bm{q}}_n},{{\bm{p}}_1},...,{{\bm{p}}_m}} {G_{{\bm{k}} + {\bm{q}}_T^{\left( n \right)} + {\bm{p}}_T^{\left( m \right)}}^{\left( 0 \right)}\left( {i{\omega _n} - \left( {m - n} \right)\Omega } \right)} 
\nonumber \\ & & \times 
\sum\limits_{j = 1}^m {{G_{n,m - 1}}\left( {{\bm{k}},i{\omega _n},{{\bm{q}}_1},...,{{\bm{q}}_n},{{\bm{p}}_1}...,\widehat {{{\bm{p}}_j}},...,{{\bm{p}}_m}} \right)} 
\nonumber \\
& & + \frac{{{\delta _{n,m}}}}{{{N^{n + m}}}}\sum\limits_{{{\bm{q}}_1},...,{{\bm{q}}_n},{{\bm{p}}_1},...,{{\bm{p}}_m}} {G_{{\bm{k}} + {\bm{q}}_T^{\left( n \right)} + {\bm{p}}_T^{\left( m \right)}}^{\left( 0 \right)}\left( {i{\omega _n} - \left( {m - n} \right)\Omega } \right)} 
\nonumber \\ & & \times 
{\delta _{{\bm{q}}_T^{\left( n \right)} + {\bm{p}}_T^{\left( n \right)},0}}\left\langle {b_{{{\bm{q}}_1}}^\dag b_{{{\bm{q}}_2}}^\dag ...b_{{{\bm{q}}_n}}^\dag {b_{ - {{\bm{p}}_1}}}{b_{ - {{\bm{p}}_2}}}...{b_{ - {{\bm{p}}_n}}}} \right\rangle. 
\eea

Now we introduce an approximation
\be
G_{{\bm{k}} + {\bm{q}}_T^{\left( n \right)} 
+ {\bm{p}}_T^{\left( m \right)}}^{\left( 0 \right)}
\left( {i{\omega _n} - \left( {m - n} \right)\Omega } \right)
\simeq G_{{\bm{k}} + {{\bm{Q}}_{n - m}}}^{\left( 0 \right)}
\left( {i{\omega _n} 
- \left( {m - n} \right)\Omega } \right).
\ee
where ${\bm Q}_n = {\bm Q}$ for $n$ odd and 
${\bm Q}_n = 0$ for $n$ even.
Applying this approximation to the equation above, we find
\bea
{g_{n,m}}\left( {{\bm{k}},i{\omega _n}} \right)
& \simeq & 
g G_{{\bm{k}} + {{\bm{Q}}_{n - m}}}^{\left( 0 \right)}\left( {i{\omega _n} - \left( {m - n} \right)\Omega } \right)
\nonumber \\ & & \times 
\left[ {\sqrt N {g_{n,m + 1}}\left( {{\bm{k}},i{\omega _n}} \right) + \sqrt N {g_{n + 1,m}}\left( {{\bm{k}},i{\omega _n}} \right) + \frac{m}{{\sqrt N }}{g_{n,m - 1}}\left( {{\bm{k}},i{\omega _n}} \right)} \right]
\nonumber \\
& & + \frac{1}{N}\frac{{n!}}{{{{\left[ {{e^{\beta \Omega }} - 1} \right]}^n}}}G_{{\bm{k}} + {{\bm{Q}}_{n - m}}}^{\left( 0 \right)}\left( {i{\omega _n}} \right){\delta _{n,m}}.
\label{eq:gnm_rec}
\eea
Here, we have used that
\be
\left\langle {b_{{{\bm{q}}_1}}^\dag b_{{{\bm{q}}_2}}^\dag ...b_{{{\bm{q}}_n}}^\dag {b_{ - {{\bm{p}}_1}}}{b_{ - {{\bm{p}}_2}}}...{b_{ - {{\bm{p}}_n}}}} \right\rangle  \simeq n!{\left[ {{n_B}\left( \Omega  \right)} \right]^n},
\label{eq:bbb}
\ee
where the effect of the conduction electrons 
is neglected in computing this quantity.

It is instructive to see the lowest order term.
Within $O(g^2)$, the Green's function is given by
\bea
G\left( {{\bm{k}},i{\omega _n}} \right) 
\simeq \frac{1}{{i{\omega _n} - {\varepsilon _{\bm{k}}} - {g^2}\left[ {\frac{{{n_B}\left( \Omega  \right) + 1}}{{i{\omega _n} 
- {\varepsilon _{ \bm{k}+{\bm{Q}} }} - \Omega}} + \frac{{{n_B}\left( \Omega  \right)}}{{i{\omega _n} 
- {\varepsilon _{\bm{k}+{\bm{Q}}}} + \Omega}}} \right]}}.
\label{eq:G2nd}
\eea
\end{widetext}
This is a standard result obtained for a fermion-boson 
coupled system.\cite{mahan2013many}

\subsection{
\label{sec:lowT}
Dilute Magnon Approximation
}
Now we consider low-temperatures,
where $T\ll \Omega$, and the number of 
excited magnons is small.
In this case, Eq.~(\ref{eq:bbb}) with $n>0$ can be neglected.
For $n>0$, 
${g_{n,m}}\left( {{\bm{k}},i{\omega _n}} \right)$
includes the scattering process of the conduction electron
absorbing $n$ magnons.
However, this kind of processes can be ignored because $n_B(\Omega)\ll 1$.

Under this approximation, Eq.~(\ref{eq:gnm_rec})
is simplified to the following form:
\bea
\lefteqn{
{g_m}\left( {{\bm{k}},i{\omega _n}} \right) 
}
\nonumber \\
& \equiv & {g_{0,m}}\left( {{\bm{k}},i{\omega _n}} \right)
\nonumber \\
& \simeq & 
gG_{{\bm{k}} + {{\bm{Q}}_m}}^{\left( 0 \right)}\left( {i{\omega _n} - m\Omega } \right)
\nonumber \\
& \times & \left[ {\sqrt N {g_{m + 1}}\left( {{\bm{k}},i{\omega _n}} \right) + \frac{m}{{\sqrt N }}{g_{m - 1}}\left( {{\bm{k}},i{\omega _n}} \right)} \right],
\eea
and the Green's function is given by
\be
G\left( {{\bm{k}},i{\omega _n}} \right) = G_{\bm{k}}^{\left( 0 \right)}\left( {i{\omega _n}} \right)\left[ {1 + g\sqrt N {g_1}\left( {{\bm{k}},i{\omega _n}} \right)} \right].
\ee
Note that 
${g_0}\left( {{\bm{k}},i{\omega _n}} \right) 
= G\left( {{\bm{k}},i{\omega _n}} \right)$.

From this recursion equation, we find
the continued fraction form of the Green's function:
\be
G\left( {{\bm{k}},i{\omega _n}} \right) \simeq \frac{1}{{i{\omega _n} - {\varepsilon _{\bm{k}}} - \frac{{{g^2}}}{{i{\omega _n} - {\varepsilon _{{\bm{k}} + {\bm{Q}}}} 
- \widetilde \Omega  - \frac{{2{g^2}}}{{i{\omega _n} - {\varepsilon _{\bm{k}}} - 2\widetilde \Omega  - \frac{{3{g^2}}}{{...}}}}}}}},
\ee
with $\widetilde \Omega  = \Omega  + i\Gamma$.
The parameter $\Gamma$ is introduced to include
the magnon damping effect.\cite{Barabanov1995,Winterfeldt1999}

This Green's function is computed 
by diagonalizing
the tridiagonal matrix, $H_G$,
whose $i$, $j$ component is given by
\be
{\left( {{H_G}} \right)_{ij}} 
= \varepsilon _{\bm{k}}^{\left(  +  \right)}{\delta _{ij}} 
+ {\left( {{H_m}} \right)_{ij}},
\ee
with 
\be
{\left( {{H_m}} \right)_{jj}} 
= {\left( { - 1} \right)^{j - 1}}
\varepsilon _{\bm{k}}^{\left(  -  \right)} 
+ \left( {j - 1} \right)\widetilde \Omega, 
\ee
and
\be
{\left( {{H_m}} \right)_{j,j + 1}} 
= {\left( {{H_m}} \right)_{j + 1,j}} = \sqrt j g.
\ee
Here, 
$\varepsilon _{\bm{k}}^{\left(  \pm  \right)} 
= \left( {{\varepsilon _{\bm{k}}} 
\pm {\varepsilon _{{\bm{k}} + {\bm{Q}}}}} \right)/2$.
The other components of $H_m$ are zero.
The poles of the Green's function are obtained
from the eigenvalues of $H_G$
and their weights are computed from the eigenvectors of $H_G$.
We compute the spectral function by analytic continuation,
$i\omega_n \rightarrow \omega + i \delta$.
We take $\delta$ as a parameter for
the broadening of the bare conduction electron spectrum.
In general, the parameters, $g$, $\Omega$, and $\delta$
are temperature dependent.
Investigation of their temperature dependence requires
more elaborate calculations,
which is not considered in this paper.

\section{
\label{sec:result}
Result}
\begin{figure}[htbp]
\includegraphics[width=0.5\textwidth]{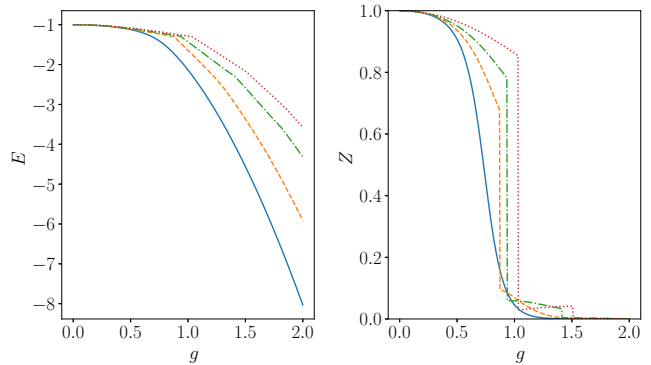}
\caption{
  \label{fig:EZ_Gamma} 
  (Color online)  
  The coupling constant $g$ dependence of 
  the magnetic polaron energy (left) 
  and the quasi-particle weight (right)
  for
  $\Gamma=0$ (solid line),
  $\Gamma=0.3$ (dashed line),
  $\Gamma=0.5$ (dash-dotted line),
  and
  $\Gamma=1.0$ (dotted line).
  Here, we take $\Omega=0.5$.
  The maximum number of the magnons is taken as
  \tm{$M=200$}.
}
\end{figure}

In Fig.~\ref{fig:EZ_Gamma}, we show
the magnetic polaron energy and the quasi-particle weight as a function of $g$.
Here, $\left( {\varepsilon _{\bm{k}}}
+ {\varepsilon _{\bm{k}+\bm{Q}}} \right)/2$
is taken as the origin of the energy.
Then, the whole spectrum depends on ${\bm k}$ 
through $\left( {\varepsilon _{\bm{k}}}
- {\varepsilon _{\bm{k}+\bm{Q}}} \right)/2 \equiv \varepsilon$
with $|\varepsilon |$ being taken as the unit of energy.
We clearly see a crossover from a weak coupling
regime to a strong coupling regime
around $g\sim 0.7$ for $\Gamma=0$.
Here, we take $M=200$ for the maximum number of the magnons.
We checked that this value is sufficiently large
and the result does not change by increasing this number.
The characteristic value of $g$ for the crossover
decreases with increasing $\Gamma$
due to the damping of the magnons.

\begin{figure}[htbp]
\includegraphics[width=0.5\textwidth]{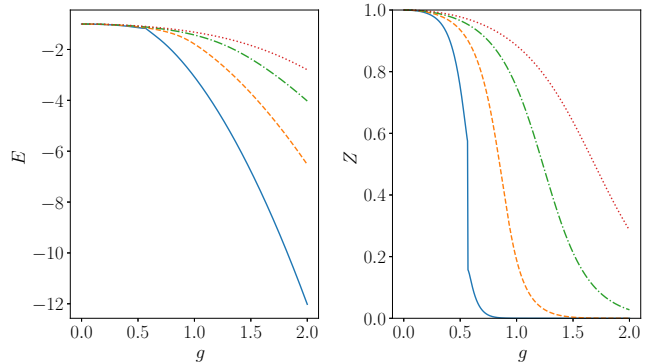}
\caption{
  \label{fig:EZ_Gamma2} 
  (Color online)
  The coupling constant $g$ dependence of 
  the magnetic polaron energy (left) 
  and the quasi-particle weight (right)
  with $\Gamma=0.1$ and \tm{$M=200$} 
  for
  $\Omega=0.3$ (solid line),
  $\Omega=0.6$ (dashed line),
  $\Omega=1.0$ (dash-dotted line),
  and
  $\Omega=1.5$ (dotted line).
}
\end{figure}

A similar behavior is observed when we change the value of $\Omega$.
In Fig.~\ref{fig:EZ_Gamma2}, we show
the magnetic polaron energy and 
the quasi-particle weight
as a function of $g$
for different values of $\Omega$.
The effect of the coupling between the conduction electron
and the magnons is suppressed by increasing $\Omega$.

\begin{figure}[htbp]
\includegraphics[width=0.5\textwidth]{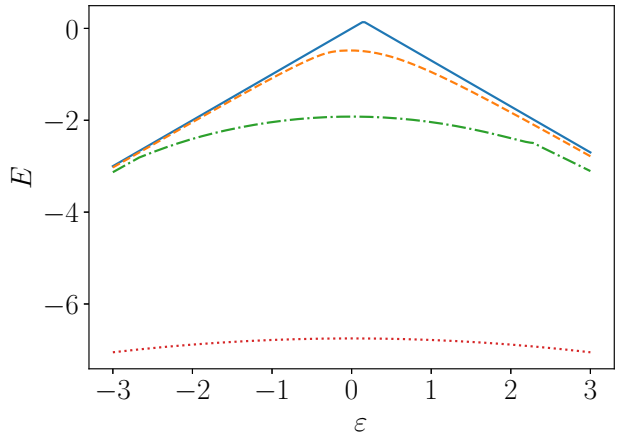}
\caption{
  \label{fig:PolaronDisp} 
  (Color online)
  The magnetic polaron energy as a function of $\varepsilon$
  with \tm{$M=200$}
  and $\Gamma=0.1$
  for
  $g=0$ (solid line),
  $g=0.4$ (dashed line),
  $g=0.8$ (dash-dotted line),
  and
  $g=1.5$ (dotted line).
}
\end{figure}

The small magnetic polaron behavior is clearly
seen when we plot the magnetic polaron energy
as a function of $\varepsilon$ as shown in Fig.~\ref{fig:PolaronDisp}.
The band width of the original conduction electrons
is reduced as we increase $g$.
This behavior is associated with 
the crossover from a large magnetic polaron 
to a small magnetic polaron.
We obtain almost flat dispersion for $g>1$.
We may expect that the conduction electrons 
localize in the presence of impurities
in the small magnetic polaron regime.
The situation is similar to self-trapping phenomena
in the polaron physics.\cite{Stoneham2007}

\begin{figure*}
\includegraphics[width=0.9\textwidth]{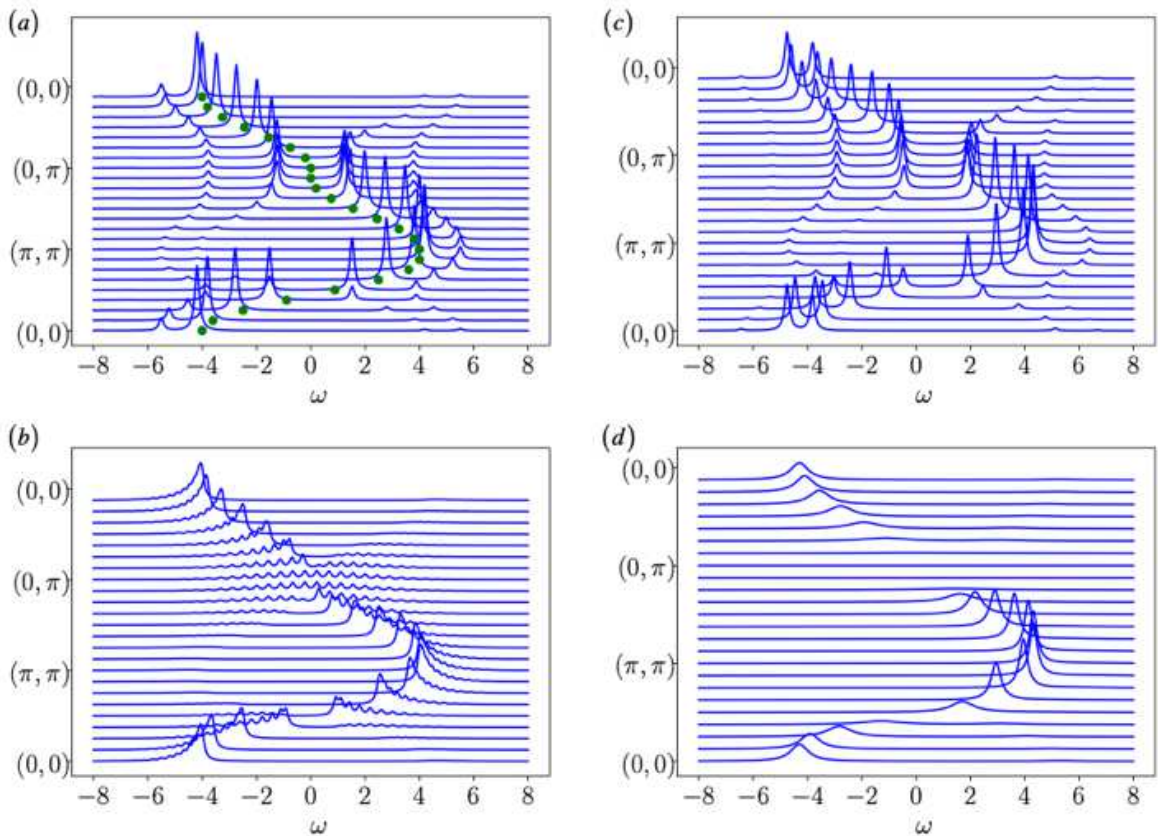}
\caption{
  \label{fig:A} 
  (Color online)
  Spectral function along symmetry directions with $g=2$
  for 
  (a) $\Omega=0$, $\Gamma=0$, and $M=6$,
  (b) $\Omega=0$, $\Gamma=0$, and $M=200$,
  (c) $\Omega=0.4$, $\Gamma=0$, and $M=6$,
  (d) $\Omega=0.4$, $\Gamma=0.3$, and $M=200$.
  The hopping parameters are $t=1$ (unit of the energy), 
  $t_1=0$, and $t_2=0$.
  The dots in (a) represent the values of 
  ${\varepsilon _{\bm{k}}}$.
}
\end{figure*}

The importance of including a sufficient number
of magnons is clarified by investigating
the spectral function.
In Fig.~\ref{fig:A} we plot the spectral function
along symmetry directions.
Here, the hopping parameter $t$ is taken as the unit of energy.
We infer the properties of the AF long-range ordered state
by taking the $\Omega \rightarrow 0$ limit.
For the case of small $M$,
that is, $M=2,4,6$, for instance,
we find a four-peak structure
along the line from $(0,0)$ to $(0,\pi)$
and near $(\pi/2,\pi/2)$
both for the AF case, $\Omega=0$, 
and for the paramagnetic case, $\Omega \neq 0$.
Similar structure was obtained in 
the cellular dynamical mean-field theory.\cite{Kyung2006,Tremblay2006,Macridin2006}
\tm{
  In cellular dynamical mean-field theory, 
  two sharp bands appear inside the Mott-Hubbard bands.
  These two bands are absent in the single-site 
  dynamical mean-field theory.
  Therefore, the appearance of these two bands
  is associated with the short-range spin correlations.
  Our Green's function approach reproduces 
  this feature for small $M$.
  But these two bands disappear
  for large enough $M$
  as shown in Fig.~\ref{fig:A}(b) and (d).
  Around $(0,\pi)$ and $(\pi/2,\pi/2)$,
  there are no quasi-particle bands.
  The band energy at these wave vectors
  is degenerate with that with 
  the wave vectors shifted by $(\pi,\pi)$.
  Therefore, many magnons are involved in 
  the dynamics of the quasi-particles 
  at these wave vectors.
  The disappearance of these two bands
  is due to the long-range correlation effect,
  which is not included either in a 
  cellular dynamical mean-field theory with 
  small cluster sizes
  or quantum Monte Carlo simulations not
  carried out at sufficiently lower temperatures.
  In the latter, the correlation length is limited to be short-ranged.
}

\tm{
  We also note that the difference between 
  $(0,\pi)$ and $(\pi/2,\pi/2)$.
  As clearly seen from Fig.~\ref{fig:A}(d),
  the scattering effect due to magnons is much significant
  at $(0,\pi)$ than at $(\pi/2,\pi/2)$.
  The difference arises from the density of states:
  The density of states is large at $(0,\pi)$
  compared to that at $(\pi/2,\pi/2)$.
  This makes the difference between the hot spot, $(0,\pi)$,
  and the cold spot, $(\pi/2,\pi/2)$.
  The lifetime of the quasiparticles is short
  for the former compared to the latter.
}

\begin{figure}[htbp]
  {
    \includegraphics[width=0.5\textwidth]{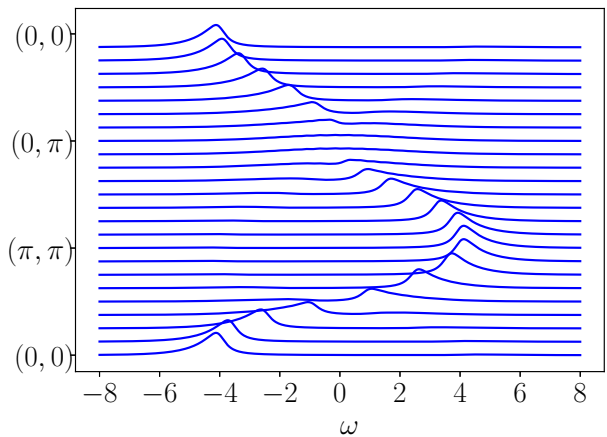}
    \caption{
      \label{fig:AFd} 
      (Color online)
      Spectral function along symmetry directions
      for the AF long-range ordered state ($\Omega=0$)  with
      the broadening $\delta=0.3$ for the bare conduction electron spectrum.
      The other parameters are the same as those of Fig.~\ref{fig:A}(b).
    }
  }
\end{figure}

We also note that the spectra shown in
Fig.~\ref{fig:A}(b) with $\Gamma=0$ and $M=200$
consist of a number of peaks with small separation.
It was pointed out that 
the broad spectra observed in the ARPES measurements
can be based on Franck-Condon broadening
in the undoped cuprates.\cite{Shen2004}
Similar broad spectra were obtained
in a diagrammatic Monte Carlo simulation
in Ref.~\onlinecite{MishchenkoNagaosa2004}
based on the $t$-$J$ model with electron-phonon coupling.
Here, a similar structure is obtained
from the electron-magnon coupling.
For a single hole doped case, we expect
that the damping effect is large in the hole dynamics.
A realistic spectrum is obtained by taking a moderate
value for the broadening of the bare conduction electron spectrum, $\delta$,
as shown in Fig.~\ref{fig:AFd}.
We note that the broad spectra,
with the width of the order of $2J$,
are associated with
the electron-magnon coupling in the strong coupling regime.

\begin{figure}[htbp]
\includegraphics[width=0.5\textwidth]{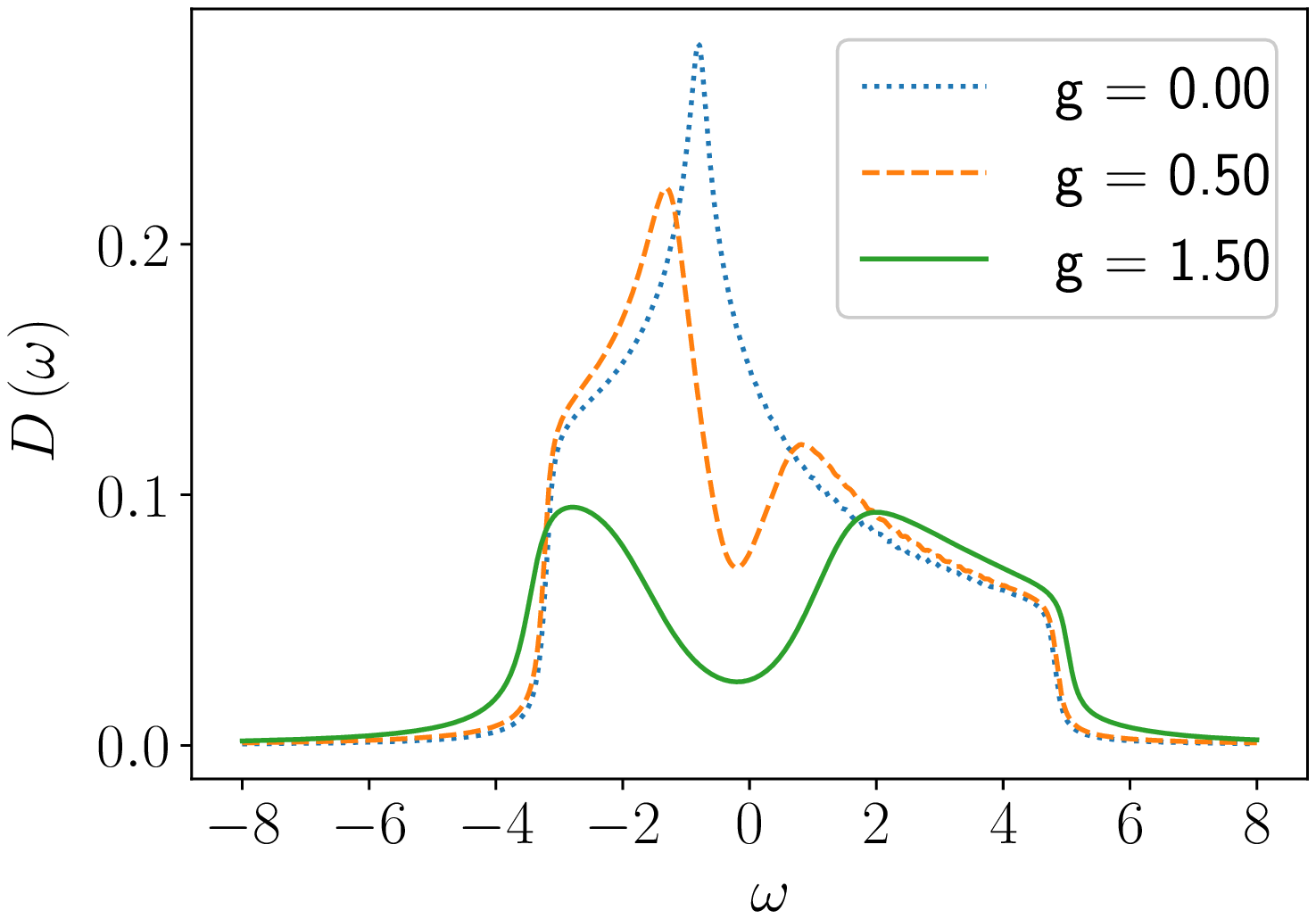}
\caption{
  \label{fig:M200} 
  (Color online)
  The density of states for different values of $g$.
  The other parameters are $\Omega=0.4$, $t_1=-0.2$, $t_2=0$, 
  $\Gamma=0.3$, $M=200$, and $\delta=0.1$.
}
\end{figure}

\begin{figure}[htbp]
\includegraphics[width=0.5\textwidth]{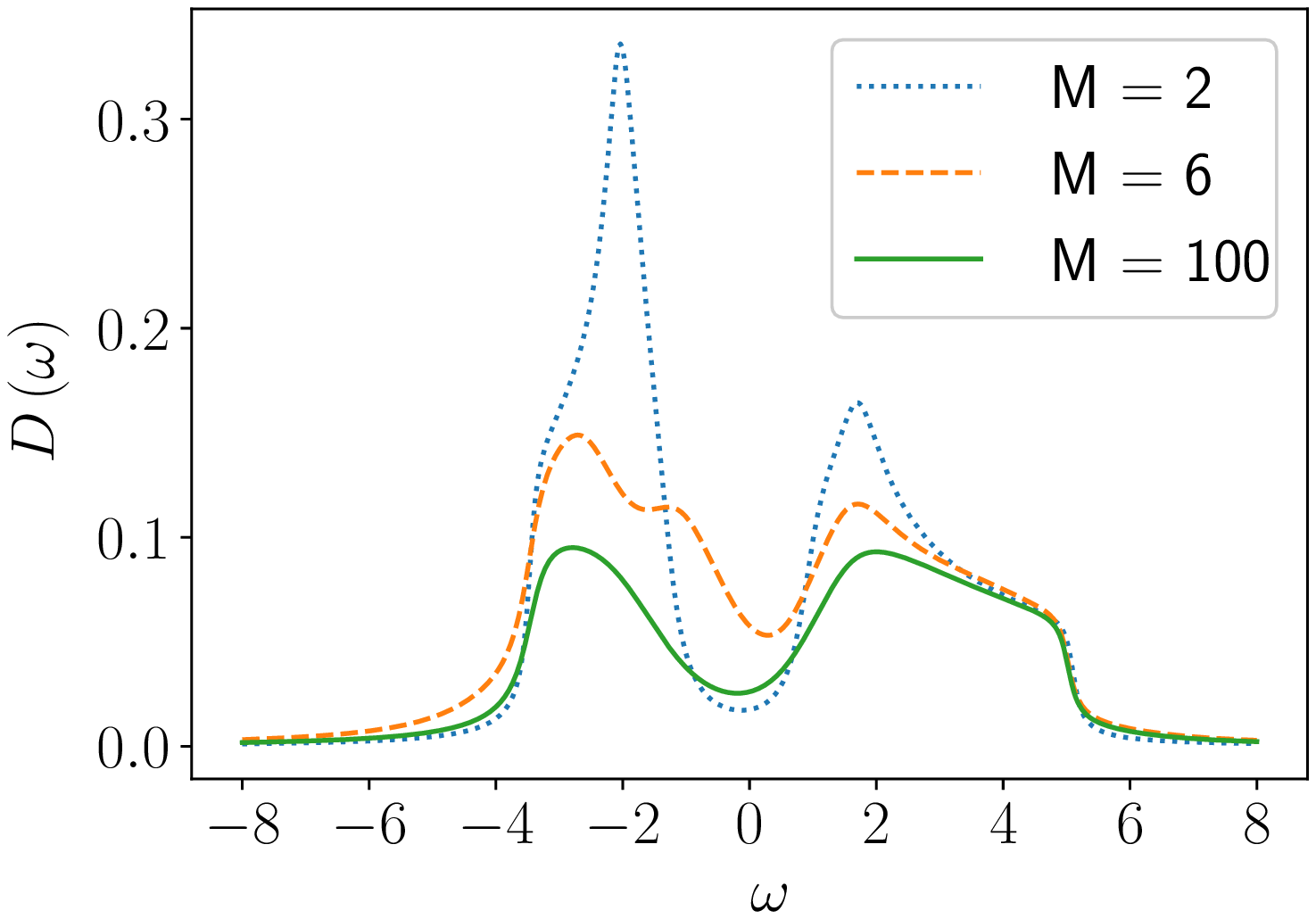}
\caption{
  \label{fig:Mdep} 
  (Color online)
  The density of states for different values of $M$.
  The other parameters are
  $\Omega=0.4$, $t_1=-0.2$, $g=1.5$, $\Gamma=0.3$, 
  $\delta=0.1$.
}
\end{figure}

The density of states is shown in 
Fig.~\ref{fig:M200} for different values of $g$
and in Fig.~\ref{fig:Mdep} for different values of $M$.
The asymmetry is associated with non-zero value of $t_1$.
We clearly see a pseudogap like behavior for large $g$.
It should be noted that
there is no Hubbard band structure
because we do not include the strong correlation
effect associated with the on-site Coulomb repulsion
between the conduction electrons.
The two-broad-peak structure for $\omega > 0$
and $\omega < 0$ is associated with
the short-range AF correlation.
Similar features,
which are well separated from the Hubbard bands,
were observed in 
the cellular dynamical mean-field theory.\cite{Kyung2006,Tremblay2006,Macridin2006}
We note that there is some change in the density of states
as we increase $M$.
We note that in these figures the total weights
decrease as we increase $g$ or $M$ because of 
the damping of the magnons.



\section{
\label{sec:summary}
Summary
}
To summarize, we have proposed a 
simple model describing 
a coupling between the AF short-range correlation
and the conduction electrons.
The model captures the essential features
of the strongly correlated electron system
except for the Hubbard band.
Applying a strong coupling analysis,
we find a crossover from a large magnetic polaron
to a small magnetic polaron.
If we limit the number of magnons, $M$, small,
we obtain the result similar to the cellular 
dynamical mean field theory.
However, taking a small number for $M$ 
corresponds to restricting the momentum resolution.
By increasing $M$, some features disappear
because of the long-range AF correlation effect
but a gap like feature remains.
This analysis suggests that including 
the long-range correlation is crucial 
and restricting the momentum resolution
can lead to incorrect results.

\tm{
The strong coupling analysis given in this paper
provides supplemental information for
the powerful numerical techniques,
like extended versions of dynamical mean-field theory
and quantum Monte Carlo simulations.
In those numerical calculations,
the Mott-Hubbard bands are clearly obtained.
However, the important feature appears
in the low-energy electronic structure 
by including the short-range correlation effect.
In general, including the short-range correlation effect
is difficult in dynamical mean-field theory
and quantum Monte Carlo simulations.
For the former, one needs a large cluster size
depending on the length of the correlation length.
However, increasing the cluster size is not an easy task.
For the latter, one needs to carry out the simulation
at low temperature depending on the length of the correlation length.
However, there is a limitation arising from the notorious sign problem.
Meanwhile, the strong coupling analysis
presented in this paper has no limitation
concerning the short-range correlation effect.
The main drawback is the absence of the 
Mott-Hubbard bands at high-energy.
}

Our analysis can be applied to
understand the strong electronic correlation
effect in the cuprate high-temperature superconductors.
For the case of the parent compound,
where a photohole is introduced 
in the ARPES measurements,
the spectra become broad
due to Franck-Condon broadening
because of the strong coupling between magnons
and the conduction electron.
The situation is similar 
to the electron-phonon coupling case.\cite{MishchenkoNagaosa2004}
Although it is natural to expect that there is contribution
from the electron-phonon coupling,
the major role can be played by the strong coupling between magnons 
and the conduction electron.
As for the paramagnetic phase,
a pseudogap like behavior has been obtained
as shown in Fig.~\ref{fig:A}(c) and (d) and in Fig.~\ref{fig:M200}.

There are several points, which are left for future research.
Since we need sufficiently large number for $M$,
multiple magnons are strongly bound to a doped hole.
This suggests that a spin texture is formed around 
a doped hole.
One possibility is a skyrmion.\cite{MorinariMFS0}
Including the dynamics of the spin texture
can lead to a modification of the energy dispersion
in the parent compound and
the gap opening in the pseudogap phase.\cite{Hashimoto2014,Vishik2018}
We also note that a small magnetic polaron behavior
may play some role in recently observed charge order,
which seems to be correlated with superconductivity.\cite{Wu2011,Ghiringhelli2012,Chang2012}
In order to study this correlation effect,
we need to consider a finite number of conduction electrons.

\begin{acknowledgments}
The author thanks T. Yoshida, D. Ootsuki, and H. Yamase for helpful comments.
\end{acknowledgments}

\appendix
\section{
\label{app:AFSRO}
Short-Range AF Correlation
}
In this appendix, we review the Green's function formalism.
Suppose we consider operators $A$ and $B$
and their Matsubara Green's function
\be
{G_{AB}}\left( \tau  \right) 
=  - \left\langle 
{{T_\tau } A\left( \tau  \right) B\left( 0 \right)} 
\right\rangle  
\equiv \left\langle {A} 
\right| \left.
{B} \right\rangle_\tau.
\ee
Here, $T_{\tau}$ is the imaginary time ordering operator
and $A\left( \tau  \right) = {e^{\tau H}}A{e^{ - \tau H}}$
with $H$ being the Hamiltonian and $\tau$ the imaginary time.
The Fourier transform of this Green's function is
\be
{G_{AB}}\left( {i{\omega _n}} \right) 
= \int_0^\beta  {d\tau } {e^{i{\omega _n}\tau }}
\left[ { - \left\langle {{T_\tau }A\left( \tau  \right)
B\left( 0 \right)} \right\rangle } \right] \equiv 
\left\langle {A} 
\right| \left. 
 {B} \right\rangle _{i{\omega _n}}.
\ee
Here, $\beta=1/(k_{\rm B}T)$ with $k_{\rm B}$ the Boltzmann constant.
For the case that $A$ and $B$ are bosonic operators,
the Matsubara frequency is given by $\omega_n=2\pi n/\beta$,
with $n$ an integer.
For the case that $A$ and $B$ are fermionic operators,
the Matsubara frequency is given by $\omega_n=\pi (2n+1)/\beta$,
with $n$ an integer.
Hereafter, we consider the former case.
Taking the derivative of 
${G_{AB}}\left( {i{\omega _n}} \right)$ with respect to $\tau$,
and Fourier transforming,
we obtain
\be
i{\omega _n}\left\langle {A} 
\right| \left.
{B} \right\rangle_{i{\omega _n}}
= \left\langle {{\left[ {A,H} \right]}} 
\right| \left.
 {B} \right\rangle_{i{\omega_n}}
+ \left\langle \left[ A,B \right] \right\rangle.
\ee

In general, we need to consider the equation of motion 
for the quantity 
$\left\langle {{\left[ {A,H} \right]}} 
\right| \left.
{B} \right\rangle _{i{\omega _n}}$,
which is given by
\be
i{\omega _n}{\left\langle {{\left[ {A,H} \right]}}
\right| \left.
{B} \right\rangle _{i{\omega _n}}}
= {\left\langle {{\left[ {\left[ {A,H} \right],H} \right]}}
\right| \left.
{B} \right\rangle _{i{\omega _n}}}
+ \left\langle {\left[ {\left[ {A,H} \right],B} \right]} \right\rangle.
\ee
Again, we need to consider the equation of motion for
the first term in the right-hand side.
To obtain a closed set of equations,
we need to introduce a Tyablikov's decoupling at some point.

Now we return to the spin system and apply the formalism above.
We define the following Matsubara Green's function,
\be
{D_{ij}}\left( \tau  \right) 
=  - \left\langle {{T_\tau }S_i^ + \left( \tau  \right)S_j^ 
- \left( 0 \right)} \right\rangle
\ee
with $S_j^ \pm  = S_j^x \pm iS_j^y$.
Here,
$S_i^ + \left( \tau  \right) 
= \exp \left( {\tau {\mathcal{H}_{{\rm{spin}}}}} \right)S_i^ 
+ \exp \left( { - \tau {\mathcal{H}_{{\rm{spin}}}}} \right)$.
The equation of motion is
\be
i{\omega _n}{\left\langle {{S_i^ + }}
\right| \left.  
{{S_j^ - }} \right\rangle _{i{\omega _n}}}
 = \left\langle {{\left[ {S_i^ + ,
{\mathcal{H}_{{\rm{spin}}}}
} \right]}}
\right| \left.
{{S_j^ - }} \right\rangle _{i{\omega _n}}
 + \left\langle {\left[ {S_i^ + ,S_j^ - } \right]} \right\rangle,
\label{eq:DSiSj_1st}
\ee
where
\be
\left\langle {{S_i^ + }}
\right| \left.
{{S_j^ - }} \right\rangle _{i{\omega _n}}
= \int_0^\beta  {d\tau } {D_{ij}}\left( \tau  \right)
\exp \left( {i{\omega _n}\tau } \right),
\ee
is the Fourier transform of 
${D_{ij}}\left( \tau  \right)$
with $\omega_n$ the bosonic Matsubara frequency.

\begin{figure}[htbp]
\includegraphics[width=0.5\textwidth]{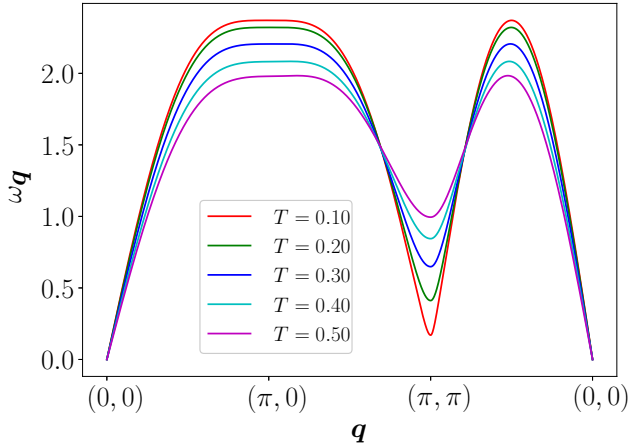}
\caption{
  \label{fig:wq} 
  (Color online)
  The magnon dispersion along symmetry directions 
  for different temperatures.
}
\end{figure}

The equation of motion for the first term in the right-hand side
of Eq.~(\ref{eq:DSiSj_1st}) is
\bea
\lefteqn{
i{\omega _n}{\left\langle {{\left[ {S_i^ + ,
{\mathcal{H}_{{\rm{spin}}}}
} \right]}}
\right| \left.
{{S_j^ - }} \right\rangle _{i{\omega _n}}} 
} \nonumber \\
&=& \left\langle {{\left[ {\left[ {S_i^ + ,
{\mathcal{H}_{{\rm{spin}}}}
} \right],
{\mathcal{H}_{{\rm{spin}}}}
} \right]}}
\right| \left.
{{S_j^ - }} \right\rangle _{i{\omega _n}}
\nonumber \\
& & + \left\langle \left[ {\left[ {S_i^ + ,
{\mathcal{H}_{{\rm{spin}}}}
} \right],S_j^ - } \right] \right\rangle.
\eea
After a tedious calculation, we obtain the explicit forms
for the two terms in the right-hand side.\cite{Kondo1972,Shimahara1991}
And then, we apply Tyablikov's decoupling
\cite{Kondo1972,Shimahara1991}
and obtain a closed form of the equations
for the Green's function.
The magnon dispersion $\omega_{\bm q}$ with the gap at $(\pi,\pi)$
can be computed as shown in 
Fig.~\ref{fig:wq} by solving the self-consistent equations numerically.

\section{
\label{app:q-dependence}
\tm{
The Effect of the Wave Vector ${\bm q}$ Dependence
of $g_{\bm q}$ and $\omega_{\bm q}$
}
}
\tm{
In the strong coupling analysis of the self-energy,
we neglected the ${\bm q}$ dependence
of $g_{\bm q}$ and $\omega_{\bm q}$ in the main text.
It is natural to ask how 
the ${\bm q}$ dependence of these parameters
affects the result.
Unfortunately, we are unaware of how one can examine this point:
If we consider $n$-th order of the self-energy,
we need to carry out ${\bm q}$ summation $n$ times.
Furthermore, just computing the self-energy up to a finite order 
is not enough to investigate the strong coupling effect.
However, it is possible to examine 
the ${\bm q}$ dependence in the weak coupling regime.
In this appendix, we consider
the effect of the ${\bm q}$ dependence of 
$g_{\bm q}$ and $\omega_{\bm q}$
within the second-order perturbation theory.
}

\tm{
From the Green's function approach given in Appendix~\ref{app:AFSRO},
we find that the dispersion of the magnon excitation
is given by
\be
{\omega _{\bf{q}}} = \frac{\Omega }{{2b}}
\sqrt {\left( {1 - {\gamma _{\bf{q}}}} \right)
\left( {1 + 2{b^2} + {\gamma _{\bf{q}}}} \right)}.
\ee
Using ${\omega _{\bf{q}}}$, 
the ${\bm q}$ dependent coupling, $g_{\bm q}$,
is expressed as
\be
{g_{\bf{q}}} = \frac{{1 - {\gamma _{\bf{q}}}}}{{2{\omega _{\bf{q}}}}}\Omega g.
\ee
The parameters $b$ and $\Omega$ are determined from the 
self-consistent calculation in the AF Heisenberg model
but here we take them as parameters and set $b=0.1$.
In this case, $g_q^2$ exhibits a sharp peak
at ${\bm q}=(\pi,\pi)$.
}

\begin{figure}
\includegraphics[width=0.5\textwidth]{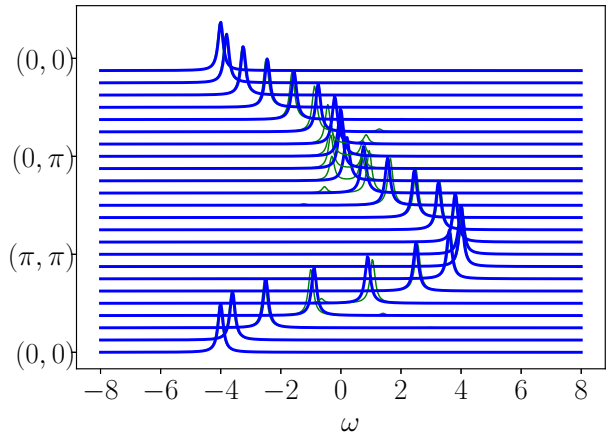}
\caption{
  \label{fig:q-dep} 
  \tm{
  (Color online)
  The spectral function along symmetry directions
  computed by Eq.~(\ref{eq:G2nd_qdep}).
  The values of the parameters are,
  $g=0.5$, $\Omega=0.4$, $b=0.1$.
  Thick blue lines are computed by Eq.~(\ref{eq:G2nd_qdep})
  while thin green lines are computed by 
  replacing $g_{\bm q}$ with $g$ and 
  $\omega_{\bm q}$ with $\Omega$
  and not taking the ${\bm q}$ summation.
}
}
\end{figure}

\tm{
Replacing $\Omega$ with $\omega_{\bm q}$
and $g$ with $g_{\bm q}$,
the conduction electron Green's function 
with the second order self-energy 
is given by
\be
{G_{\bm{k}}}^{(2)}\left( {i{\omega _n}} \right) 
= \frac{1}{{i{\omega _n} - {\varepsilon _{\bm{k}}} 
- \frac{1}{{N}}\sum\limits_{\bm{q}} 
{\frac{{{g_{\bm{q}}^2}}}{{i{\omega _n} 
- {\varepsilon _{{\bm{k}} + {\bm{q}}}} - {\omega _{\bm{q}}}}}} }}.
\label{eq:G2nd_qdep}
\ee
Figure \ref{fig:q-dep} shows the spectral function
along symmetry directions.
Here, we set $g=0.5$, which is not in the strong coupling regime,
to focus on the wave vector ${\bm q}$ dependence.
We see that there are no discernible
changes in the spectral function
except for around $(0,\pi)$.
This is simply understood as the result
of the average with respect to ${\bm q}$.
From the behavior of $g_q^2$, the dominant contribution
comes from ${\bm q}=(\pi,\pi)$.
But in taking the average in the self-energy,
this contribution is smeared out.
Therefore, the spectral function with the wave vector ${\bm q}$
dependence is almost equivalent to the non-interacting case.
}

\tm{
We note that the result shown in Fig.~\ref{fig:q-dep}
is based on the second-order self-energy.
We may expect that there are band energy changes and modification
of the spectral weights at higher-order self-energies.
However, there can be some cancellation at higher-order terms
because of the presence of ${\bm q}$ summation
in the higher-order terms.
This point is left for future research.
}

\bibliographystyle{apsrev4-2}
\bibliography{../../../refs/lib}

\end{document}